\documentclass[preprint,12pt]{elsarticle}

\usepackage{amssymb}
\usepackage{amsmath}
\usepackage{bm}
\usepackage{url}

\usepackage[dvipsnames]{xcolor}

\journal{Journal of the Mechanics and Physics of Solids}

\begin{document}
\begin{frontmatter}

\title{Kinetic Friction of Structurally Superlubric 2D Material Interfaces}
\author[inst1,inst2]{Jin Wang}
\author[inst4]{Ming Ma}
\author[inst1,inst2,inst3]{Erio Tosatti \corref{cor1}}
\ead{tosatti@sissa.it}
\cortext[cor1]{Corresponding author}

\affiliation[inst1]{organization={International School for Advanced Studies (SISSA)},
            addressline={Via Bonomea 265}, 
            city={Trieste},
            postcode={34136}, 
            country={Italy}}
\affiliation[inst2]{organization={International Centre for Theoretical Physics (ICTP)},
            addressline={Strada Costiera 11}, 
            city={Trieste},
            postcode={34151}, 
            country={Italy}}
\affiliation[inst3]{organization={CNR-IOM, Consiglio Nazionale delle Ricerche - Istituto Officina dei Materiali, c/o SISSA},
            addressline={Via Bonomea 265}, 
            city={Trieste},
            postcode={34136}, 
            country={Italy}}
\affiliation[inst4]{organization={State Key Laboratory of Tribology, Department of Mechanical Engineering},
            addressline={Tsinghua University}, 
            city={Beijing},
            postcode={100084}, 
            country={China}}

\begin{abstract}
The ultra-low kinetic friction $F_\mathrm{k}$ of 2D structurally superlubric interfaces, connected with the fast motion of the incommensurate moir\'e pattern, is often invoked for its linear increase with velocity $v_0$ and area $A$, but never seriously addressed and calculated so far. Here we do 
that, exemplifying with a twisted graphene layer sliding on top of bulk graphite -- a demonstration case that could easily be generalized to other systems.
Neglecting quantum effects and assuming a classical Langevin dynamics, we derive friction expressions valid in two temperature regimes.
At low temperatures the nonzero sliding friction velocity derivative
$\mathrm{d}F_\mathrm{k}/\mathrm{d} v_0$
is shown by Adelman-Doll-Kantorovich type approximations to be equivalent to that of a bilayer whose substrate is affected by an analytically derived effective damping parameter, replacing the semi-infinite substrate. At high temperatures, friction grows proportional to temperature as analytically required by fluctuation-dissipation.
The theory is validated by non-equilibrium molecular dynamics simulations with different contact areas, velocities, twist angles and temperatures.
Using $6^\circ$-twisted graphene on Bernal graphite as a prototype we find a shear stress of measurable magnitude,
from $25$~kPa at low temperature to $260$~kPa at room temperature, yet only at high sliding velocities such as $100$~m/s.
However, it will linearly drop many orders of magnitude below measurable values at common experimental velocities such as $1~\mu \mathrm{m/s}$,
a factor $10^{-8}$  lower.
The low but not ultra-low ``engineering superlubric" friction measured in existing experiments should therefore be attributed to defects and/or edges, whose contribution surpasses by far the negligible moir\'e contribution.

\end{abstract}



\end{frontmatter}


\section{Introduction}
Structural superlubricity (SSL) is the phenomenon where the low temperature static friction $F_\mathrm{s}$-- the minimal force required to start sliding -- of a theoretically infinite crystal-crystal contact is identically zero \cite{Shinjo.surfsci.1993,Martin.phystoday.2018,Vanossi.rmp.2013}. In that state the  kinetic friction $F_\mathrm{k}$-- the force required to maintain steady-state sliding with velocity $v_0$ -- is nonzero, and believed to grow proportionally to $v_0$ as $v_0 \to 0$. Incommensurate interfaces between 2D materials (incommensurability guaranteeing perfect cancellation of lateral forces) are good candidates for SSL, owing to their high in-plane stiffness, weak interlayer interaction, and small corrugations \cite{Song.small.2020,Vanossi.nc.2020,Yuan.nanotech.2022}. 
Since the first ``experimental verification" in nanoscale graphite flakes at room temperature \cite{Dienwiebel.prl.2004}, experiments on SSL have been promoted in various directions with diverse foci, including large scales \cite{Liu.prl.2012,Zhang.natnano.2013}, high speed \cite{Yang.prl.2013,Peng.pnas.2020}, low temperature \cite{Liu.App.Mat.2020}, and hetero-structures \cite{Dietzel.prl.2013,Song.NatureMaterials.2018,Liao.natmater.2022,Panizon.Nanoscale.2022, Cao.prx.2022}.
Besides their intrinsic physical interest, superlubricity phenomena are of potential relevance in mechanical engineering, data storage and aerospace \cite{Hod.nature.2018}.\\

A first striking dichotomy one encounters in literature
is that while most experimental work reports kinetic friction,
theoretical studies of SSL concentrate on static friction \cite{deWijn.prb.2012,Koren.prb.moire.2016}, which is distinctly different \cite{Wang.rmp.2022}.
As a result, kinetic friction, despite the abundance of data and its all-important energy dissipation significance, is only modestly modeled and analytically understood \cite{Ma.prl.2015,Wang.nanolett.2019,Mandelli.prl.2019}. Early studies of incommensurate Frenkel-Kontorova (FK) chains showed that the kinetic friction of this one dimensional SSL system, whose static friction is nominally zero, may turn especially large when exciting phonons whose wavevector is the 1D equivalent of the inverse moir\'e wavelength \cite{Consoli.PhysRevLett.2000}, a finding likely to carry over to 2D.
Although general formulations of kinetic friction are long established \cite{Adelman.jcp.1976,Kantorovich.prb1.2008}, there are currently no explicit calculations of $F_\mathrm{k}$ for realistic SSL systems at a generic velocity. In particular, predicting kinetic friction from the structural and physical properties of  incommensurate structurally lubric 2D material interface, where the  swift ``surfing" motion of the moir\'e pattern is the only source of the extremely small dissipation, stands as an open question of basic significance. \\

Here, using a twisted graphene layer on semi-infinite Bernal graphite as a demonstrative example, we derive analytical expressions of the kinetic friction force of structurally superlubric sliding friction expected to be valid without any fitting or fudge parameters. Following a preliminary part, where the geometry and mechanical bases of the 2D twisted material interface are described (Sect.~2), our approach consists of four steps. In the first step, starting from a Langevin equation of motion for a bilayer, consisting of a sliding layer and a substrate layer mimicking a semi-infinite bulk, we analytically connect friction to the dissipation incurred by the moving moir\'e pattern (Sect.~3).
Besides its correct proportionality to velocity and to area, the result is also proportional to the substrate layer's Langevin damping constant $\zeta$  -- alas an arbitrary and uncontrolled parameter. In the second step (Sect.~4), we specialize the Adelman-Doll-Kantorovich basic surface scattering  approach \cite{Adelman.jcp.1976,Kantorovich.prb1.2008} to calculate precisely the Langevin damping $\zeta$, making that no longer an uncontrolled parameter. 
In the third step, we consider in Sect.~5 the opposite limit of high temperatures where friction is analytically determined by fluctuation-dissipation, and where a very useful heuristic interpolation from low to high temperatures is proposed.
In Sect.~6 we validate these theoretical results by direct comparison with non-equilibrium molecular dynamics (MD) sliding simulations, first of a twisted graphene bilayer, and eventually of the twisted monolayer sliding on multilayer graphite, where the multilayer extrapolation technique of Benassi \textit{et al.} \cite{Benassi.prb.2010} is implemented to obtain the true, damping-independent kinetic friction.
A discussion concludes the paper in Sect.~7.\\

As one could physically anticipate, the true, defect-and-edge-free structurally superlubric friction of a 2D material interface, whose frictional stress is due exclusively to the gossamer-like moir\'e flight, is generally irrelevant. That is, it is of such small magnitude to be practically unaccessible to measurement, at least in the commonest low-velocity sliding experiments.
Probably for this reason the pure moir\'e kinetic friction is often invoked but never calculated, let alone subjected to a theory-simulation comparison and discussion.
That however does not diminish its importance. Calculating the SSL kinetic friction will provide an element of clarity, conceptual and practical. Conceptually, it is instructive to estimate the tiny friction elicited by moving discommensurations, quasiparticle-like entities that fly and dissipate energy in Stokes-like fashion inside a viscous medium. Practically, these results are required to substantiate (or to deny) the hunch that much of the supposedly superlubric kinetic friction  data reported in literature, of very measurable magnitude and not of linear velocity and area dependence, must in reality be of different origin, usually connected with stick-slip caused by a multiplicity of defects and edges rather than to the Stokes flight of the perfect moir\'e \cite{Wang.rmp.2022}. As it will turn out, the hunch is vastly confirmed.

\begin{figure}[ht!]
\centering
\includegraphics[width=\linewidth]{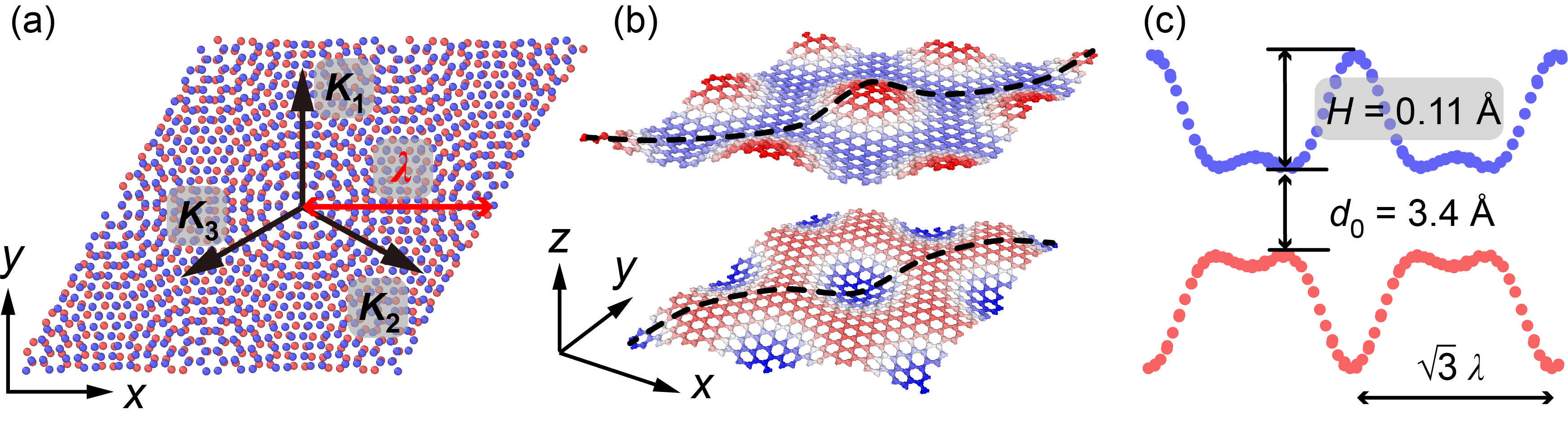}
\caption{Structure of twisted bilayer graphene. (a) Unrelaxed structure. The lattice constant $\lambda$ and the direction of reciprocal vectors $\textbf{\textit{K}}_n$ of the moir\'e lattice are illustrated. (b) The relaxed structure ($\theta=6^\circ$), where upper and lower layers are colored according to the out-of-plane displacement. (c) The cross-section along the AA-to-2nd nearest AA direction (dashed line in b). For clarity, the out-of-plane deformation in (b) and (c) is enlarged and the interlayer spacing is reduced. The real values are marked in the figure.}
\label{fig:1}
\end{figure}

\section{Modeling the moir\'e corrugation pattern and mechanics}

A twisted graphene bilayer -- lower layer the ``substrate", upper layer the ``slider" -- is shown in Fig.~\ref{fig:1}, depicted at a twist angle $\theta=6^\circ$ as an example. The relaxed structure (b) and (c)  shows the large out-of-plane deformation of both layers concentrated at discommensurations -- the most important tribological characteristics, stemming from the low bending stiffness of membrane-like 2D materials \cite{Wang.prl.2019,Han.NatMater.2020}.

The potential field experienced by an atom at point $ (\textbf{\textit{r}}, d) $ in the sliding layer can be modeled as
\begin{equation}
    U(\textbf{\textit{r}},d)= \frac{2U_0 e^{\alpha (1-d/d_0)}}{9} \sum_{n=1}^3 \cos(\textbf{\textit{K}}_n \cdot \textbf{\textit{r}})+ [-\varepsilon + k (d-d_0)^2]
    \label{eq:1}
\end{equation}
The first term on the
right-hand side describes the moir\'e-modulated corrugation potential, where $d$ is the interlayer distance at $\textbf{\textit{r}}$, $d_0$ is the equilibrium distance for the unrelaxed bilayer, $U_0$ is the sliding energy barrier when $d=d_0$, $\alpha$ represents the decay rate of that barrier as $d$ is increased, and $\textbf{\textit{K}}_n$ is the reciprocal vector of the moir\'e superlattice ($n=1, 2, 3$). Specifically, $\textbf{\textit{K}}_n=\textbf{\textit{k}}_n-\textbf{\textit{q}}_n$,
where $\textbf{\textit{k}}_n$ and $\textbf{\textit{q}}_n$ are the reciprocal vectors of the substrate and the upper slider.
With rotation tensor $\textbf{\textit{R}}(\theta)$, they are connected by $\textbf{\textit{q}}_n=\textbf{\textit{R}}(\theta) \cdot \textbf{\textit{k}}_n$, where $\theta$ is the twist angle.
The second term represents interlayer adhesion, with $\varepsilon$ the average adhesion energy per atom, and $k$ its $d$-dependence. Using for pure convenience a 6-12 Lennard-Jones-type interlayer interaction of depth parameter $\varepsilon$ and range $d_0$ for our parameter representation, then $k=36 \varepsilon /d_0^2$. A potential expression such as Eq.(1), generic but perfectly adequate for relatively small twist angles, was also validated in previous work \cite{Wang.nanolett.2019}.
Besides neglecting the very high harmonics connected with non-sinusoidality of the potential exerted by each layer, the additional assumptions made, both in agreement with simulations, are
\begin{enumerate}
    \item The in-plane displacement of each atom away from the layer's center-of-mass is zero. That is 
    justified by the large in-plane stiffness of 2D materials.
    \item The maximum out-of-plane displacement, i.e., the moir\'e height modulation amplitude $H$ is much smaller than the interlayer distance $d_0$. In graphene for example, $H\leq (d_\mathrm{AA}-d_\mathrm{AB})/2$, whence $H/d_0 \leq 0.1/3.4 \sim 3\%$ \cite{Wijk.2dmater.2015}.
\end{enumerate}

Based on Eq.~(1), the $z$-direction force field exerted  by the substrate on the slider $F_z(\textbf{\textit{r}},d) = -\partial U/\partial d $ can be written as
\begin{equation}
    F_z(\textbf{\textit{r}},d)= \frac{2\alpha U_0 e^{\alpha (1-d/d_0)}}{9 d_0} \sum_{n=1}^3 \cos(\textbf{\textit{K}}_n \cdot \textbf{\textit{r}}) + 2k(d_0 - d)
    \label{eq:2}
\end{equation}
Since $|d-d_0|\ll d_0$, the contribution from the second (adhesion) term is negligible. Thus the monolayer out-of-plane deformation field can be approximated by

\begin{equation}
    w(\textbf{\textit{r}})= \frac{2\alpha U_0 e^{\alpha (1-d/d_0)}}{9 k' d_0} \sum_{n=1}^3 \cos(\textbf{\textit{K}}_n \cdot \textbf{\textit{r}})= \frac{2H}{9} \sum_{n=1}^3 \cos(\textbf{\textit{K}}_n \cdot \textbf{\textit{r}})
    \label{eq:3}
\end{equation}
where $k'$ is an effective out-of-plane stiffness and $H$ is the moir\'e corrugation height.
Thus, the interlayer distance between two layers (of opposite corrugation) is
\begin{equation}
    d(\textbf{\textit{r}})= \langle d \rangle + \frac{4H}{9} \sum_{n=1}^3 \cos(\textbf{\textit{K}}_n \cdot \textbf{\textit{r}})
    \label{eq:4}
\end{equation}
where $\langle d \rangle$ is the equilibrium distance for the relaxed bilayer. For convenience, we define $\delta=d_0-\langle d \rangle$ ($0\leq \delta \ll d_0$).

In the next step, an explicit expression for $H$ will be obtained by the conservation relationship between the interfacial adhesive work and the bending deformation energy during the structural relaxation (from unrelaxed flat to relaxed corrugated structure).

\subsection{Adhesive energy}
The adhesive energy per atom of the corrugated graphene sheet is
\begin{equation}
    E_\mathrm{adh}=\frac{1}{A_\mathrm{m}} \int_\mathrm{moire} U(\textbf{\textit{r}},d) \,\mathrm{d}A
    \label{eq:5}
\end{equation}
The integral is on 
the moiré cell, of area $ A_\mathrm{m} =\sqrt{3} \lambda^2/2$,
where the moir\'e lattice constant is $\lambda=\sqrt{3}a/\sqrt{2-2 \cos{\theta}}$ \cite{Hermann.jpcm.2012}, and $a$ the bond length of graphene.
Due to the sixfold angular periodicity, twist angle $\theta$ can be extended from domain $(0, \pi/6]$ to any angle.
Based on the small deformation assumption, we obtain
\begin{equation}
    E_\mathrm{adh}\approx
    -\varepsilon - \frac{4\alpha U_0 H}{27 d_0}
    +\frac{32\varepsilon H^2}{3 d_0^2}
    \label{eq:6}
\end{equation}
The first term is the adhesive energy for flat bilayers, the remaining ones describe the perturbation introduced by the out-of-plane moir\'e corrugation.

\subsection{Bending energy}
The (per atom)
bending energy cost of the monolayer moir\'e corrugation is
\begin{equation}
        E_\mathrm{bend}=\frac{\kappa}{2N(1-\nu^2)} \int_\mathrm{moire} \{(\frac{\partial^2 w}{\partial x^2}+\frac{\partial^2 w}{\partial y^2})^2 +
        2(1-\nu)[(\frac{\partial^2w}{\partial x \partial y})^2- \frac{\partial^2 w}{\partial x^2}\frac{\partial^2 w}{\partial y^2}]\}\,\mathrm{d}A
        \label{eq:7}
\end{equation}
where $N$, $\kappa$ and $\nu$ are the atom number, bending rigidity and Poisson's ratio of the monolayer.
Substituting in Eq.~(\ref{eq:3}), the above equation simplifies to
\begin{equation}
    E_\mathrm{bend}=\frac{64\pi^4 \kappa a^2 H^2}{27\sqrt{3}(1-\nu^2)\lambda^4}
    \label{eq:8}
\end{equation}

\subsection{Bulk slider elastic energy}
Most experimental studies and applications involving sliding of 2D materials interfaces require them to be deposited or encapsulated \cite{Song.NatureMaterials.2018,Liao.natmater.2022}, rather than
freestanding. The perpendicular substrate and sliding stage elasticity also limits out-of-plane deformations. To model this effect, also needed for our successive connection with sliding on thick graphite, we attach perpendicular springs (with spring constant $k_z$) to all atoms of substrate and slider to limit their out-of-plane deformation.
The total elastic energy of monolayer graphene normalized to  the area of a single atom is
\begin{equation}
\begin{aligned}
    E_\mathrm{spring}&=\frac{1}{A_\mathrm{m}} \int_\mathrm{moire} \frac{1}{2}k_z w^2 \,\mathrm{d}A
    &=\frac{1}{27} k_z H^2
\end{aligned}
\label{eq:9}
\end{equation}

\subsection{Total potential energy}
The energy per slider atom of the model bilayer can thus be written as
\begin{equation}
\begin{aligned}
    U_\mathrm{tot}&=2\cdot (E_\mathrm{bend}+E_\mathrm{spring}+E_\mathrm{adh})\\
    &= \frac{128\pi^4 \kappa a^2 H^2}{27\sqrt{3}(1-\nu^2)\lambda^4}
    +\frac{2}{27}k_z H^2 - 2\varepsilon
    -\frac{8\alpha U_0 H}{27 d_0}
    +\frac{64\varepsilon H^2}{3 d_0^2}
\end{aligned}
\label{eq:10}
\end{equation}
where the prefactor 2 represents two layers. The right-hand side can be regarded as a quadratic function of $H$, $U_\mathrm{tot}=AH^2+BH+C$, where
\begin{equation}
\begin{aligned}
    A&=\frac{128\pi^4 a^2 \kappa}{27\sqrt{3}(1-\nu^2)\lambda^4}
    +\frac{2}{27} k_z
    +\frac{64\varepsilon}{3 d_0^2}\\
    B&=-\frac{8 \alpha U_0}{27 d_0}\\
    C&= -2 \varepsilon\\
\end{aligned}
\label{eq:11}
\end{equation}
Thus, the (real) moir\'e height $H$ of the slider corresponding to the minimum total potential energy is
\begin{equation}
    H=-\frac{B}{2A}=\frac{\alpha U_0}
    {\frac{32\pi^4 \kappa a^2 d_0}{\sqrt{3}(1-\nu^2) \lambda^4}
    +\frac{k_z d_0}{2}
    +\frac{144\varepsilon}{d_0}}
\label{eq:12}
\end{equation}
Here, the only twist angle-dependent parameter is the in-plane moir\'e size $\lambda(\theta)$. At large twist angles, $\lambda$ is small and so does the corresponding corrugation.
At $\theta \to 0$, $\lambda$ grows and eventually diverges as $\theta^{-1}$, whence $H$ saturates to $\frac{\alpha U_0}{k_z d_0/2 + 144\varepsilon/d_0}$.

\begin{figure}[ht!]
\centering
\includegraphics[width=\linewidth]{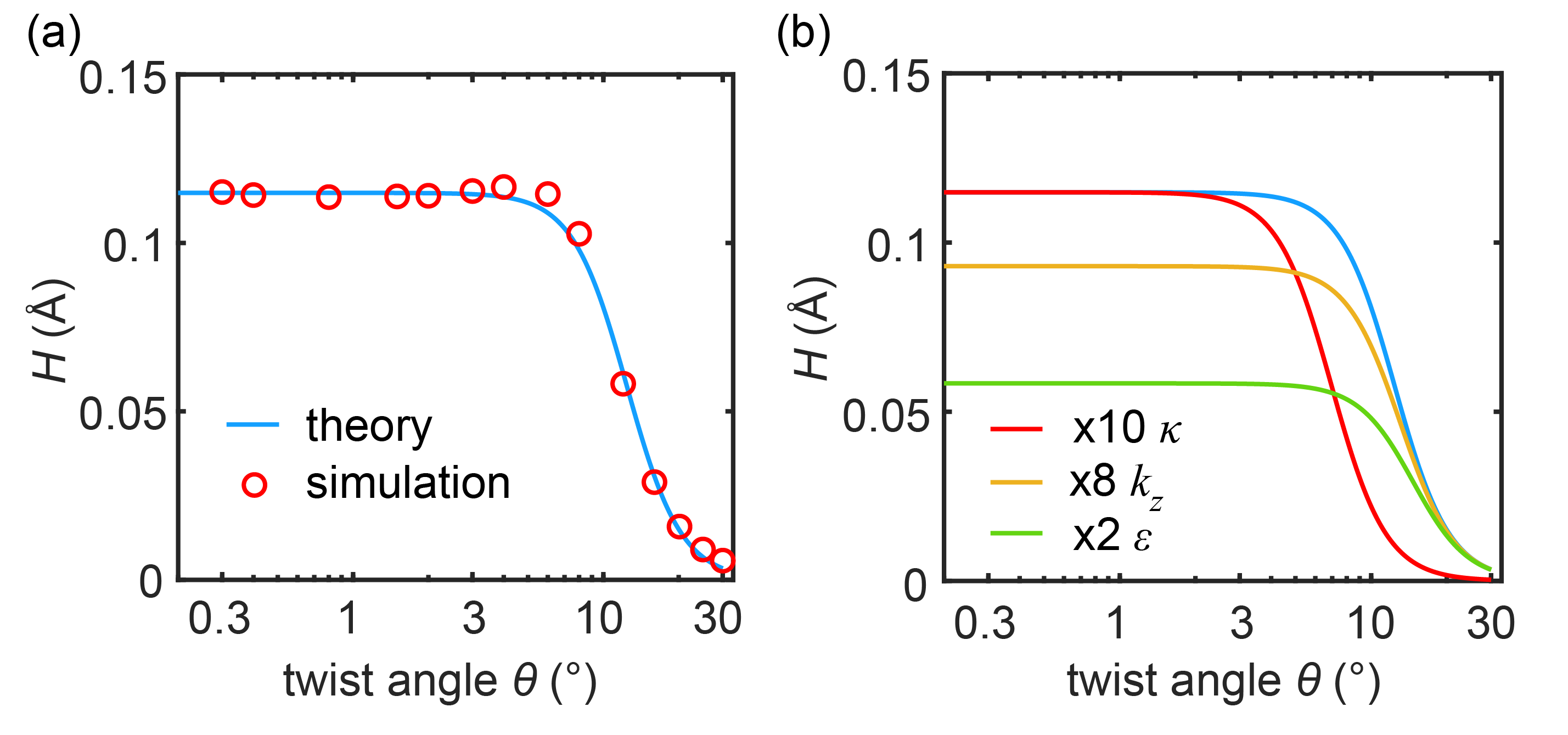}
\caption{Dependence of moir\'e corrugation amplitude on the twist angle in a graphene bilayer.
(a) Comparison of the simulation (red) and theory (blue).
The levelling off below $\sim 5^\circ$ reflects the saturation of the corrugation magnitude to its maximum extent
$(d_\mathrm{AA} - d_\mathrm{AB})/2 \sim 0.11~\mathrm{\AA}$.
(b) Theoretical corrugation for system with 10 times larger bending stiffness (red), 8 times larger supporting stiffness (yellow), and 2 times larger interlayer adhesion energy (green).}
\label{fig:2}
\end{figure}

We underline that (Eq. 12) does not contain fudge or unknown quantities, all parameters being determined by the mechanical or structural properties of the material. For twisted bilayer graphene, the interlayer parameters are: sliding energy barrier $U_0=14$~meV/atom,  decay rate $\alpha \approx 8.8$, interlayer distance $d_0=3.4$~\AA, interlayer attraction parameter $\varepsilon=24.5$~meV/atom \cite{Ouyang.nanolett.2018}. The intralayer parameters are: bending stiffness and Poisson's ratio 
$\kappa=1.4$~eV and $\nu=0.2$ \cite{Lu.JphysD.2009}, and bond length $a=1.42$~\AA \cite{Brenner.2002}. The $z$-direction supporting spring stiffness $k_z$, which reproduces the moiré height in twisted bulk graphite, is $k_z=0.33$~N/m. (That for a freestanding bilayer, would of course be $k_z=0$).\\

A direct comparison between our analytical moir\'e corrugation $H$ and molecular simulation results is shown in Fig.~\ref{fig:2}a (details for the simulation given in SI). The comparison confirms that our theoretical model is quantitatively applicable to a wide range of twist angles, from 0.3 to 30 degrees
(and of course beyond 30 degrees, owing to sixfold angular periodicity).
It is worth noting that the sinusoidal deformation assumption is not strictly applicable to the system with a twist angle
$\theta \lesssim 3^\circ$, where the moiré deformation strongly
deviates from sinusoidal \cite{Zhang.JMPS.2018, Wang.rmp.2022}, and the in-plane deformation becomes non-negligible \cite{Kazmierczak.NatMater.2021}. Nevertheless, since the saturated out-of-plane corrugation at small twist angles originates from reduced interlayer attraction in the so-called AA regions, the in-plane deformation has a negligible effect, and our simplified model still provides a good estimate for $H$.

The result (Eq. 12) can also describe the corrugation magnitude of 2D materials other than twisted graphene. For different materials/constraints, e.g., larger bending stiffness $\kappa$ (for transition-metal dichalcogenides), higher $k_z$ (for adsorbed or multilayer system), stronger binding energy $\varepsilon$, the corrugation height and its relation to the twist angle is predicted and shown in Fig.~\ref{fig:2}b.
It can be seen from the figure that higher bending stiffness reduces the moir\'e corrugation at large twist angles, while larger $k_z$ and $\varepsilon$ reduce the corrugation at small twists.

\section{Kinetic friction of a surfing  moir\'e: Langevin equation}

\begin{figure}[ht!]
\centering
\includegraphics[width=\linewidth]{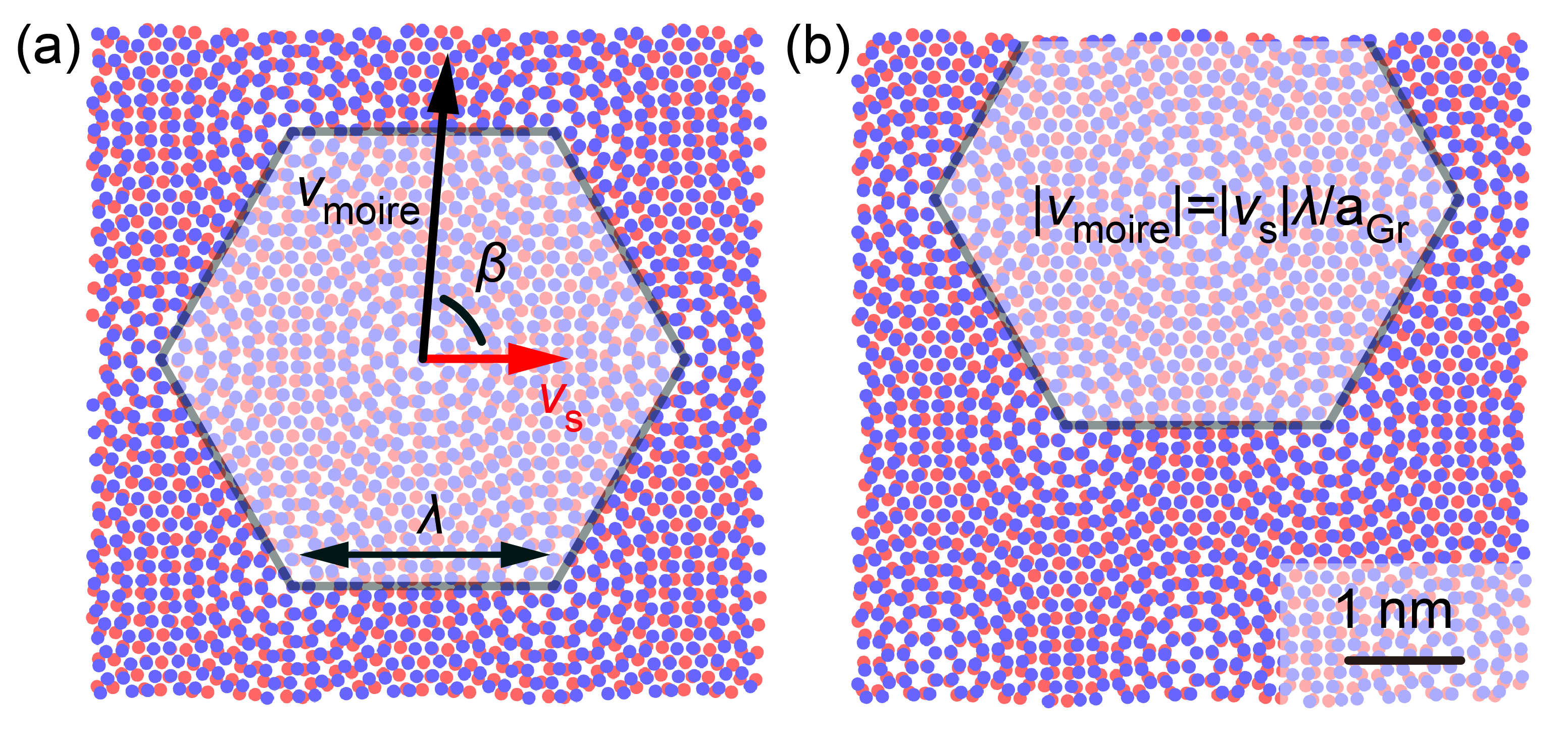}
\caption{(a) Twisted graphene. Sliding direction of the layer center-of-mass (red arrow) and of the moir\'e pattern (black arrow).
For a center-of-mass sliding distance equal to the interatomic distance $a$ ($1.42 ~\mathrm{\AA}$ along $\textbf{\textit{v}}_s$), the larger displacement of the moiré is shown in (b). Here the twist angle is $\theta=6^\circ$, $\lambda=2.35$~nm, and $\beta=87^\circ$.}
\label{fig:3}
\end{figure}

Considering a sliding process with velocity $\textbf{\textit{v}}_s$, the moir\'e out-of-plane displacement field becomes
\begin{equation}
    w(\textbf{\textit{r}})=\frac{2H}{9} \sum_{n=1}^3 \cos[\textbf{\textit{K}}_n \cdot (\textbf{\textit{r}}-\textbf{\textit{v}}_\mathrm{moire} t)]
\end{equation}
where the moir\'e surfing velocity  could be written as \cite{Hermann.jpcm.2012}:
\begin{equation}
    \textbf{\textit{v}}_\mathrm{moire}=-\frac{2}{3} \sum_{n=1}^3 \frac{(\textbf{\textit{k}}_n-\textbf{\textit{q}}_n) \otimes \textbf{\textit{q}}_n}{|\textbf{\textit{k}}_n-\textbf{\textit{q}}_n|^2} \cdot \textbf{\textit{v}}_s
\end{equation}
The magnitude of the moir\'e surfing velocity 
is related to the sliding velocity as $|\textbf{\textit{v}}_\mathrm{moire}|=v_0 \lambda /a_\mathrm{Gr}$, with $v_0=|\textbf{\textit{v}}_s|$ the sliding speed of the upper layer and $a_\mathrm{Gr}$ the layer's atomic lattice constant.
As illustrated in Fig.~\ref{fig:3}, the angle $\beta$ between the moir\'e surfing direction and the atomic sliding direction is twist angle-dependent, $\beta=(\pi-\theta)/2$. For $\theta \to 0$, moir\'e moves nearly perpendicular to the sliding direction of the upper flake.

We begin by describing the friction energy dissipated from the interface to the substrate monolayer, by the empirical -- yet fundamentally motivated -- Langevin equation, which contains a phenomenological damping coefficient $\zeta$.
For each substrate atom, 
\begin{equation}
    m_i \ddot{r}_{i, \alpha} =  -\nabla_{i, \alpha} V_\mathrm{tot} - \zeta_{\alpha} \dot{r}_{i, \alpha} + R_{i,\alpha}
\end{equation}
where $\alpha=x, y, z$, $m_i$ and $r_{i,\alpha}$ are the mass and position (along $\alpha$ direction) of $i$-th substrate atom, $V_\mathrm{tot}$ is the total potential energy of the system, $\zeta_{\alpha}$ and $R_{i,\alpha}$ are the damping coefficient and random force along direction $\alpha$.
In the Markov approximation, the random force term satisfies the fluctuation-dissipation relation
$\langle R_\alpha(t) \rangle=0$ and
$\langle R_\alpha(t)R_{\alpha'}(0) \rangle =2\zeta k_{\rm B}T \delta_{\alpha \alpha'} \delta(t)$.
Aiming first at the energy dissipation from the interface to the substrate, we assume zero temperature (thus the thermal noise term $R=0$).
The influence of finite temperature and the neglect of quantum effects will be discussed in Sect.~5.

At steady state sliding, the kinetic friction of the system satisfies power conservation
\begin{equation}
    F_\mathrm{k} v_0 = \sum_{\alpha}^{x,y,z} \zeta_{\alpha} \sum_{i=1}^N m_i \langle v_{i, \alpha}^2 \rangle
\end{equation}
where the left hand side represents the input power, the right hand side the dissipated power (each $\langle v_{i, \alpha}^2 \rangle$ being proportional to 
$v_0^2$ at $T=0$), where $N$ is the number of slider atoms.
Assuming for a 2D material bilayer a highly anisotropic damping coefficient, $\zeta_z\gg\zeta_x \sim \zeta_y$ (the reason for such assumption will be further addressed in the next section), this can be simplified to
\begin{equation}
\begin{aligned}
    F_\mathrm{k} &= \frac{\zeta_{z}}{v_0} \sum_{i=1}^N m_i \langle v_{i, z}^2 \rangle\\
    &= \frac{m \zeta_{z}}{v_0 A_\mathrm{C}} \int_\mathrm{moire} \langle v_z^2 \rangle \,\mathrm{d}A\\
    &= \frac{m \zeta_{z}}{v_0 A_\mathrm{C} \tau} \int_\mathrm{moire} \int_0^{\tau} v_z^2 \,\mathrm{d}t \,\mathrm{d}A
\end{aligned}
\end{equation}
where $A_\mathrm{C} = 3 \sqrt{3} a^2/4$ is the area per atom.
Based on Eq.~(13), the out-of-plane velocity is $v_z=\mathrm{d}w/\mathrm{d}t$, and Eq.~(17) can be further simplified to
\begin{equation}
    F_\mathrm{k}=\frac{c_1 N m \zeta_z H^2 v_0}{a_\mathrm{Gr}^2}
\end{equation}
where $c_1$ is a prefactor which depends on the out-of-plane moir\'e structural corrugation shape. For the sinusoidal structure (Eq. 3) -- a good approximation for $\theta \gtrsim 3^{\circ}$ systems, $c_1=16\pi^2/81$.
The result (Eq.~18) shows that friction of a 2D superlubric slider is proportional to the atom number $N$, therefore to the slider's contact area $A=3\sqrt{3} N a^2/4$;
proportional to the sliding velocity $v_0$ (i.e., it is viscous);
proportional to the square of moir\'e corrugation $H(\theta)$; and finally, proportional to the damping $\zeta_z$.
This is as far as a traditional Langevin formulation takes us.

\section{Fundamental derivation of damping parameter}

The analytical Eq.~(18) appears to solve the problem.
Yet, the presence in the result of the so far arbitrary damping coefficient $\zeta$ is far from satisfactory -- in real life there is no damper. Simply, frictional phonons propagate away from the interface and never come back. This situation, described by many authors
\cite{Adelman.jcp.1976, Consoli.PhysRevLett.2000, Kantorovich.prb1.2008,Benassi.prb.2010,Krylov.2014,Park.ChemRev.2014}, is physically clear but still needs a solution analytically connecting the effective bilayer damping to the real damping-free system.
Another way out is to do away with damping. For example, the multilayer extrapolation technique of Ref. \cite{Benassi.prb.2010} implies that application of a Langevin damping
$\zeta_{N_L}$, $N_L$ layers away from the sliding interface will yield the correct friction for {\it any} $\zeta_{N_L}$ in the $ N_L \to \infty$ limit.
A welcome feature of that approach is also that for any finite $N_L$ the bottom dissipation parameter $\zeta_{N_L}$ can be {\it{variationally optimized}}.
On the other hand, linear response, single-phonon friction (see e.g., \cite{Panizon.prb.2018}), free from any arbitrary damping parameters and therefore conceptually more satisfactory, generally yields a Born approximation friction formula whose applicability does not include the low velocity limit and whose result is dimension dependent. That suggests that  viscous friction, the velocity-linear friction generically expected in SSL should result from multi-phonon processes, beyond Born perturbative theory.

\begin{figure}[ht!]
\centering
\includegraphics[width=\linewidth]{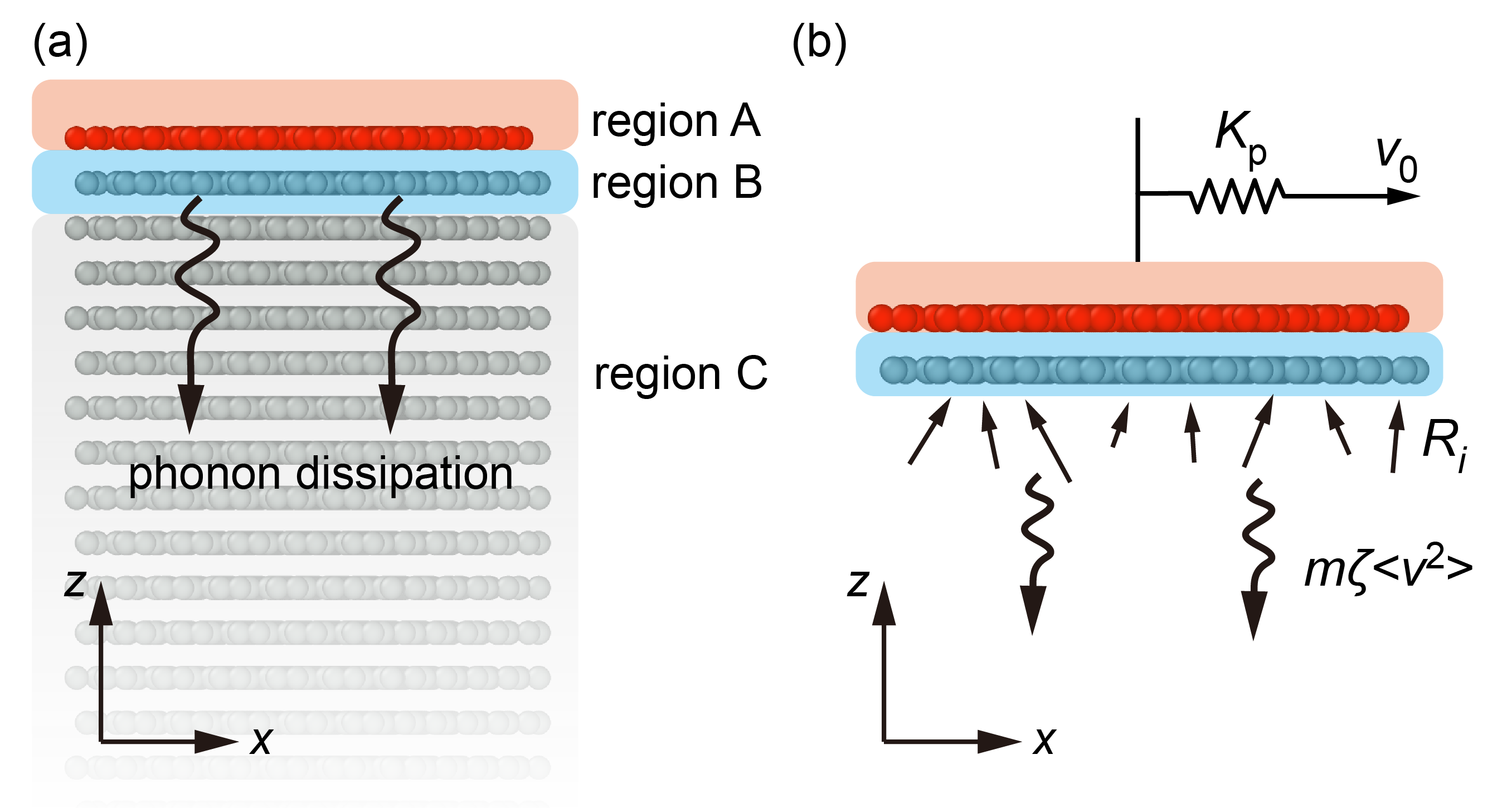}
\caption{Schematic diagram of SSL system and energy dissipation. (a) The (real) multilayer system with phonon dissipation. (b) The reduced effective bilayer system with random force $R_i$ and phonon dissipation described by the effective damping $\zeta$. The slider (region A),
topmost layer of the substrate (region B) and half-infinite remaining substrate (region C) are colored red, blue, and gray respectively.}
\label{fig:4}
\end{figure}

An effective, microscopically generated and analytically defined damping coefficient $\zeta$ can be obtained from basic principles of surface scattering, as formulated long ago \cite{Adelman.jcp.1976, Kantorovich.prb1.2008} for a harmonic system.  Consider a model 2D material friction geometry as illustrated in Fig.~\ref{fig:4}a, where region A is the slider, B is the interfacial layer of the substrate, and C is a (half-infinite) substrate.
The generalized Langevin equation for the atom in region B could 
be described by:
\begin{equation}
    m_i \ddot{r}_i = f_i - \int_0^{t} \Gamma(t,\tau) \dot{r}_i \,\mathrm{d}\tau + R_i 
\end{equation}
where $m_i$ and $r_i$ is the mass and the position of $i$-th atom, $f_i$ represents the interaction between atom-$i$ and the rest of the system, $R_i$ is considered as the random force due to the motion of atoms in region C, and the integral term describes the frictional force with the friction kernel
\begin{equation}
\begin{aligned}
    \Gamma(t,\tau)&=\beta \langle R_{i}(t) R_{i}^{\dagger}(\tau)  \rangle \\
    &= m_i D_\mathrm{BC}(t) \Pi_\mathrm{CC}(t-\tau) D_\mathrm{CB}(\tau)
\end{aligned}
\end{equation}
here $\beta=1/k_\mathrm{B}T$, $D$ is the dynamical matrix, $\Pi_\mathrm{CC}=\sum_{\lambda} e_{\lambda} \otimes e_{\lambda}^{\dagger} \cos(\omega_{\lambda}t)/\omega_{\lambda}^2$, where $e_{\lambda}$ and $\omega_{\lambda}$ are the eigenvector and eigenvalue of the substrate (C region). By neglecting the long-range interactions and in a  Markov approximation \cite{Kantorovich.prb2.2008}
(checked to be reasonable for our systems with short relaxation times from fs to ps,
a simplified damping term $m_i \zeta \dot{r_i}$ may replace the friction term in Eq.~(19), yielding the effective damping coefficient
\begin{equation}
    \zeta_{\alpha}=\int_0^{t} D_\mathrm{BC} \Pi_\mathrm{CC}(\tau) D_\mathrm{CB} \,\mathrm{d}\tau
\end{equation}
Since direct interaction between atoms in region B and atoms deep down in region C is negligible, the
effective frictional kernel can be reduced to a first-neighbour one, so that the above formula can be simplified as
\begin{equation}
\begin{aligned}
    \zeta_{\alpha} &\simeq \frac{|\Phi_{12}^{\alpha \alpha}|^2}{m^2} \int_0^{t} \Pi_\mathrm{CC}(\tau) \,\mathrm{d}\tau
\end{aligned}
\end{equation}
where $\Phi_{12}^{\alpha \alpha}=\frac{\partial^2 V}{\partial r_{1,\alpha} \partial r_{2,\alpha}}$ is the force constant between atom 1 (in B) and 2 (in C) along $\alpha$ direction (which could be estimated from its definition or equivalently, from the elastic constant. See SI for details), and $m$ is the mass of atom in region B.
For a half-infinite isotropic substrate, this further simplifies by using
\begin{equation}
\begin{aligned}
    \Pi_\mathrm{CC}(\tau) &\simeq \int_0^{\omega_\mathrm{D}} \rho(\omega) \frac{\cos(\omega \tau)}{\omega^2} \mathrm{d}\omega\\
    &= \frac{3}{\omega_\mathrm{D}^2} \frac{\sin(\omega_\mathrm{D} \tau)}{\omega_\mathrm{D} \tau}
\end{aligned}
\end{equation}
where $\rho(\omega)=3\omega^2/\omega_\mathrm{D}^3$ is the phonon density of states and $\omega_\mathrm{D}$ is the Debye frequency.
Thus, for $t\to \infty$, Eq.~(21) finally becomes
\begin{equation}
    \zeta_{\alpha} \simeq \frac{3\pi}{2\omega_\mathrm{D}^3} \frac{|\Phi_{12}^{\alpha \alpha}|^2}{m^2}
\end{equation}

This Eq. (24), similar to earlier single-molecule \cite{Persson.prb.1999} and bulk solid \cite{Kantorovich.prb1.2008} formulas, is our desired result for the damping coefficient, whose insertion into Eq.~(18) of previous section should lead to the friction of an SSL 2D material interface, approximate but now free of arbitrary parameters. Simple as it is, it is controlled by two fundamental quantities, both connected to the lattice dynamics of the substrate. The first is the denominator $\omega_\mathrm{D}^3$, indicating that a softer substrate
will cause a much higher viscous sliding friction. This is actually a very general result, also obtained earlier for surface vibrating molecules \cite{Persson.prb.1999}.
The second is the numerator $|\Phi_{12}^{\alpha \alpha}|^2$, a dynamical matrix term  measuring how effective the first-substrate layer transmits a friction-generated vertical vibration to the second (now Bernal commensurate) layer, the second to the third, and so on. Again as in the vibrating molecule theory this is
proportional to the fourth power of the sliding-induced $z$-oscillation frequency and therefore to the second power of the bulk phonon density of states -- a higher density of states corresponds to a higher decay rate of the interface-generated phonon, which reflects as a higher friction.\\

At the cost of ruining its simplicity, Eq.~(24) could in principle be improved by taking into account dissipation due to $(x, y)$ polarizations, vibrational anisotropy, anharmonicity, and quantum effects.  As regards including all phonon polarizations, it is physically clear and proven by many simulations \cite{Song.NatureMaterials.2018,Mandelli.prl.2019, Wang.rmp.2022}
that the main contribution to the friction of a bilayer should be due to the part of Eq.~(21) from the 2D material substrate vibrations whose $|\Phi_{12}^{\alpha \alpha}|^2$ is vertically polarized, whose $\omega_\mathrm{D}$ is the softest. To test the second point, we actually extended our estimate to an anisotropic Debye model, more appropriate for layered materials. 
That done, results yield damping coefficients of the same order of magnitude (see SI for details). Anharmonicity will only play a role at high temperatures; a regime where, as will be discussed later, a totally different approach is called for, since thermal fluctuations exceed moir\'e corrugation $H$. At the opposite extreme low temperature limit, below a temperature $T_q$ to be introduced below, quantum effects will in fact become important but will require a treatment that is beyond our present scopes. \\

Within the present approximations, whose range of validity we have thus qualified, it is interesting to look at actual orders of magnitude predicted for moir\'e friction. Considering the interlayer interaction of graphene, we estimate that
$\Phi_{12}^{zz}=2.7$~N/m (neglecting in-plane terms like $|\Phi_{12}^{xx}|^2$ that are much smaller).
Inserting the $z$-Debye frequency of graphite $\omega_\mathrm{D}=1.2\times 10^{14}$~rad/s \cite{Krumhansl.jcp.1953}, we obtain an effective damping coefficient $\zeta_z=0.05~\mathrm{ps}^{-1}$.
Even if omission of anharmonicity inevitably entails a slight underestimate -- as we shall see later by comparison with true values from simulations -- this damping coefficient is considerably smaller than empirical values currently used in 2D materials simulations, e.g., $4.5~\mathrm{ps}^{-1}$ \cite{Song.NatureMaterials.2018,Wang.nanolett.2019,Mandelli.prl.2019}, or $1~\mathrm{ps}^{-1}$ \cite{Ouyang.nanolett.2018,Gao.nc.2021}, $2~\mathrm{ps}^{-1}$ \cite{Gigli.2dmater.2017}.
Such large empirical values may have been adopted as conveniently similar to $10^3$ simulation time steps (the latter typically $\sim 1$~fs),
or adjusted by requiring independence of friction on damping,
or chosen just so as to yield a friction that is comparable to experiments. That kind of choice is particularly alarming  because the typical simulation and experimental velocities differ by six orders of magnitude, and true superlubric friction should be, and we confirm it is, linear with velocity.
Contrary to that, the experimental 2D friction generally depends logarithmically and not linearly upon velocity \cite{Wang.epl.2019,Song.NatureMaterials.2018,Liu.App.Mat.2020}. All that implies that the choice of effective Langevin damping $\zeta$ should be completely reconsidered in SSL simulations, and that validation of analytical results should be sought through comparison with genuinely superlubric PBC simulations, rather than with  experiments, where friction is ``engineering superlubric" and not structurally superlubric \cite{Wang.rmp.2022}.

\section{High temperature SSL kinetic friction}

It is now necessary to connect the low temperature results presented so far, Eqs.~(18) and (24),  with realistic finite and high temperatures. Conceptually, as the temperature increases, the random thermal flexural corrugations with amplitude $\langle |H_T| \rangle \propto \sqrt{k_\mathrm{B} T}$ become larger \cite{Fasolino.NatMater.2007}, surpassing the original moir\'e corrugation $H$, eventually making it irrelevant.
In addition, and more importantly, thermal agitation and anharmonicity will gradually suppress the phonon lifetimes and mean free paths, reducing and eventually destroying the use of single phonon approximations. We can address this limiting regime in the low velocity, high temperature ``diffusive" limit through fluctuation-dissipation, predictably leading to a linear increase of friction with $T$. 
In this limit, kinetic friction is given by
\begin{equation}
    F_\mathrm{k}=\frac{1}{\mu} v_0
\end{equation}
where $\mu$ is the drift mobility, 
connected to the diffusion coefficient $D$ and temperature $T$ by Einstein's relation $\mu=D/k_\mathrm{B} T$.
The connection between the diffusion coefficient and temperature is generally Arrhenius-like, i.e., $D(T)=D_0 \exp(-E_0/k_\mathrm{B}T)$, where $D_0$ is the maximum diffusion coefficient and $E_0$ the activation energy barrier for diffusion.
Thus, we expect  the temperature dependence of kinetic friction at finite temperature and infinitesimal velocity to follow linear response
\begin{equation}
    F_\mathrm{k} (T)=\frac{k_\mathrm{B} v_0}{D_0} T
\end{equation}

To substantiate this result, we must still specify what $D_0$ actually is. One way to do that, actually applicable only in the SSL case, is to make a connection between low velocity friction, just described, and the opposite high velocity limit of ``ballistic friction" -- described for example by Guerra \textit{et al.} \cite{Guerra.NatMater.2010} -- where additional resistance to sliding arises from collisions with large thermal fluctuations causing surfing moir\'e to
dissipate more work -- such as a raft would when surfing a rough sea. 
The low and high velocity regimes are in general quite different.
That is because energy barriers dominate ordinary friction at most velocities, only becoming irrelevant at extremely low (``thermolubric") \cite{Krylov.2014,Dong.2011} and extremely high (``ballistic") velocities \cite{Guerra.NatMater.2010}.
The peculiarity of SSL systems is precisely the absence of
energy barriers against sliding, and that merges the two regimes into a single one.
In the ballistic regime, kinetic friction can be described by replacing the moir\'e out-of-plane corrugation with thermal fluctuations:
\begin{equation}
    F_\mathrm{k} \simeq
     \frac{N m \zeta_z \langle H_T^2 \rangle v_0}{a_\mathrm{Gr}^2}
    =\frac{c_2 N m \zeta_z v_0 k_\mathrm{B} T}{\kappa}
\end{equation}
where $c_2$ is a dimensionless prefactor of order 1 which reflects the proportionality between temperature and mean square vertical fluctuations.
Even if straightforward to formulate in a harmonic approximation, the actual calculation of $c_2$ for our target geometry -- a twisted/incommensurate monolayer on top of a substrate monolayer or semi-infinite bulk-- implies summing contributions from a cumbersome variety and number of modes. In addition, a harmonic description of frictional phonons would be wrong at high temperatures. In place of that calculation we therefore simply use the $c_2$ value independently obtained by an equilibrium MD simulation.
(For a twisted and $k_z $ harnessed graphene bilayer with $\theta=6^{\circ}$, for example, the actual value is $c_2\approx2.8$).
By equating two expressions, we find that $D_0 \sim \kappa(N m \zeta_z)^{-1}$. (Note once again that the atom number $N$ is proportional to the contact area $A$ as it should).

The overall picture of the viscous kinetic friction
$F_\mathrm{k} = v_0 (\mathrm{d}F_\mathrm{k} / \mathrm{d} v_0)$
in SSL systems is thus clear. At high temperatures friction grows linearly with $T$. At lower temperatures, where the thermal fluctuation amplitude $\langle |H_q| \rangle$ is smaller than the moir\'e corrugation $H$, friction becomes temperature independent, leveling off to a value determined the out-of-plane moir\'e distortion $H$. Reflecting this physical crossover, the total kinetic friction force can be heuristically approximated by
\begin{equation}
    F_\mathrm{k} \simeq
    N m \zeta_z v_0 \sqrt{(\frac{c_1 H^2}{a_\mathrm{Gr}^2})^2
    +(\frac{c_2 k_\mathrm{B} T}{\kappa})^2}
\end{equation}
This formula, with damping $\zeta_z$ given by Eq.~(24), and with $H= H[\lambda(\theta)]$ determined by the twist angle $\theta$ through Eq.~(12), represents the main result of this paper.\\

By equating two terms in the above formula, a crossover temperature between two regimes can be estimated as
\begin{equation}
    T_c \simeq \frac{\kappa H^2}{a_\mathrm{Gr}^2 k_\mathrm{B}}
\end{equation}
For $\theta=6^\circ$ case ($H=0.12~\mathrm{\AA}$), $T_c \simeq 30$~K.

One last remark is about quantum effects that are ignored here, but that must become important at sufficiently low temperatures. Considering the equipartition energy of the classical flexural mode with the moir\'e wavelength,  and the quantum zero-point energy, one can estimate a further crossover temperature $T_q$ below which the classical theory fails \cite{Hasik.PRB.2018}:
\begin{equation}
    T_q = \frac{\hbar \omega_\mathrm{moire}}{2 k_\mathrm{B}} 
    = \frac{\hbar}{2k_\mathrm{B}} \sqrt{\frac{\kappa}{\rho_\mathrm{2D}}} (\frac{2\pi}{\lambda})^2
\end{equation}
where $\rho_\mathrm{2D}$ is the area density, and $\lambda$ is the moir\'e size. For twisted graphene  with $\theta=6^\circ$, this transition temperature is estimated to be $T_q \approx 15$~K -- below the other classical crossover $T_c$, but not much lower. This indicates that our theoretical formula (Eq. 28) should apply for $T>T_q$ (and below the $z$-Debye temperature, $T_\mathrm{D}=900$~K \cite{Krumhansl.jcp.1953}),
but it will require additional quantum modifications in the true low temperature limit.

\section{Validation by MD simulations}

Our theoretical results for structurally superlubric friction of a twisted 2D interface on top of a semi-infinite 2D crystal need to be validated. We cannot do that by comparison with experiments, because a) the experimental velocity and area dependence, generally much weaker than linear, show that their friction is not strictly structurally superlubric; b) the friction of SSL systems at their low velocity would be far too small to be measured by any available technique. We can nonetheless validate our result, by comparing our theoretical friction to a non-equilibrium MD multilayer simulation where, following Benassi \textit{et al.} \cite{Benassi.prb.2010}, we can obtain an approximately parameter-free friction. In this section, we firstly compare our analytical results (Eq.~28) with damping-based MD graphene bilayer simulations at low and high temperatures. Then, finite temperature ``realistic" friction is obtained using the parameter-free simulation to test our prediction of the damping coefficient (Eq. 24).

\begin{figure}[ht!]
\centering
\includegraphics[width=0.9\linewidth]{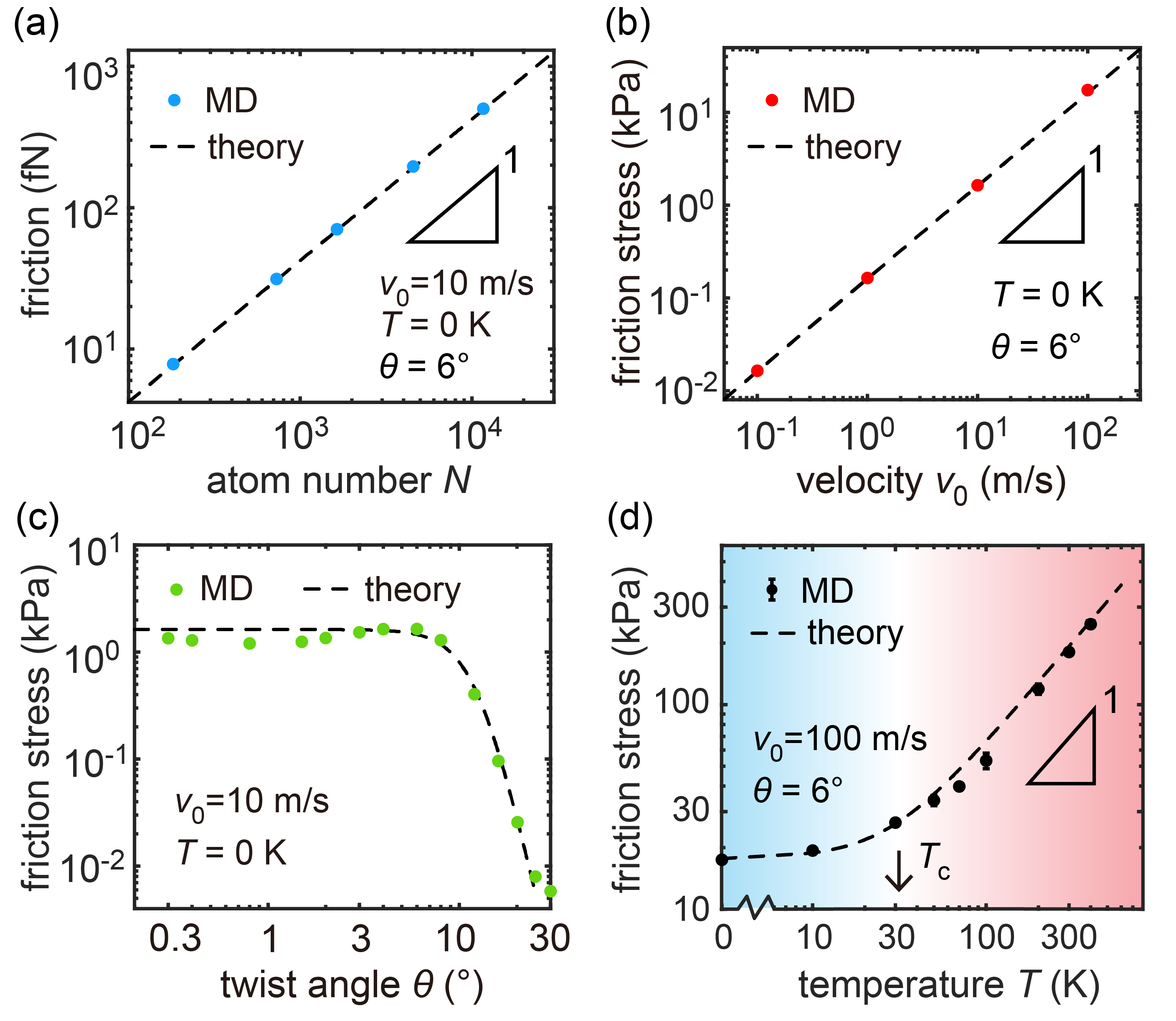}
\caption{Size (a), velocity (b), twist angle (c), and temperature (d) dependence of a twisted graphene bilayer sliding friction from MD simulation with $\zeta_z$ given by Eq.~(24) (data points) and theory of Eq.~(28) (dashed lines). The value of parameters are given in each plot.
The arrow in (d) marks the crossover temperature $T_c$, below which (blue region) friction is dominated by moir\'e corrugation and saturates to a constant, and above which (red region)
it is dominated by flexural fluctuations and grows linearly with temperature. The use of unusually high sliding velocities permits shorter simulation times, and is allowed by the completely linear dependence of friction upon sliding velocity.
}
\label{fig:5}
\end{figure}

\subsection{Damping-based bilayer simulation at low and high temperature}
The validity of the analytical low temperature damping parameter Eq. (18) can be tested by MD simulations of our model twisted graphene bilayer (Fig.~\ref{fig:4}b). Realistic model geometries with twist angles $\theta$ ranging from $0.3^\circ$ to $30^\circ$ are created by optimization of the total energy at rest with periodic boundary conditions (PBC).
The interlayer and intralayer interaction are described by registry-dependent interlayer potential (ILP) and REBO force field respectively \cite{Brenner.2002,Ouyang.nanolett.2018}.
Perpendicular springs
(with spring constant $k_z=0.33$~N/m) are attached to each atom in both substrate and slider layers to mimic the real confinement effect between the driving stage and the semi-infinite substrate. The center-of-mass of the slider is connected to a dragging spring with spring constant $K_p=100$ N/m, pulled with a constant velocity $v_0$ along $x$.
A Langevin thermostat with $T=0$~K and damping $\zeta_z$, which from Eq.~(24) is approximately $0.05~\mathrm{ps}^{-1}$, is attached to the substrate layer. The kinetic friction $F_\mathrm{k}$  is calculated as the time-averaged force experienced by the dragging spring.
Simulation results shown in Fig.~\ref{fig:5}a-c agree quantitatively with the theoretical prediction (dashed lines), confirming the linear dependence of  SSL friction upon area  ($A \propto N$), and the twist angle dependence predicted by $H[\lambda(\theta)]$ of Eq.~(12).

To validate the high temperature analytical result, we thus perform MD simulations with the same model (with twist angle $\theta=6^{\circ}$), and compare the frictional result with theory of Eq. (28). The finite temperature (thermal noise) is introduced by a Gaussian-distributed random force $R_z$ with $\langle R_z(t) R_z(0) \rangle =2\zeta_z k_{\rm B}T \delta(t)$. The damping coefficient $\zeta_z$ is given from Eq. (24).
The MD simulation results are shown in Fig.~\ref{fig:5}d (solid points).
To compare with our effective bilayer theory, we extract the dimensionless prefactor $c_2$, which describes the linear dependence of $\langle H_T^2 \rangle$ on temperature $T$, from simulations (details in SI). Substituting its value $c_2\approx2.8$ into Eq.~(28), we find excellent agreement between our theory and simulations.

\subsection{Comparison with parameter-free simulations}

\begin{figure}[ht!]
\centering
\includegraphics[width=0.9\linewidth]{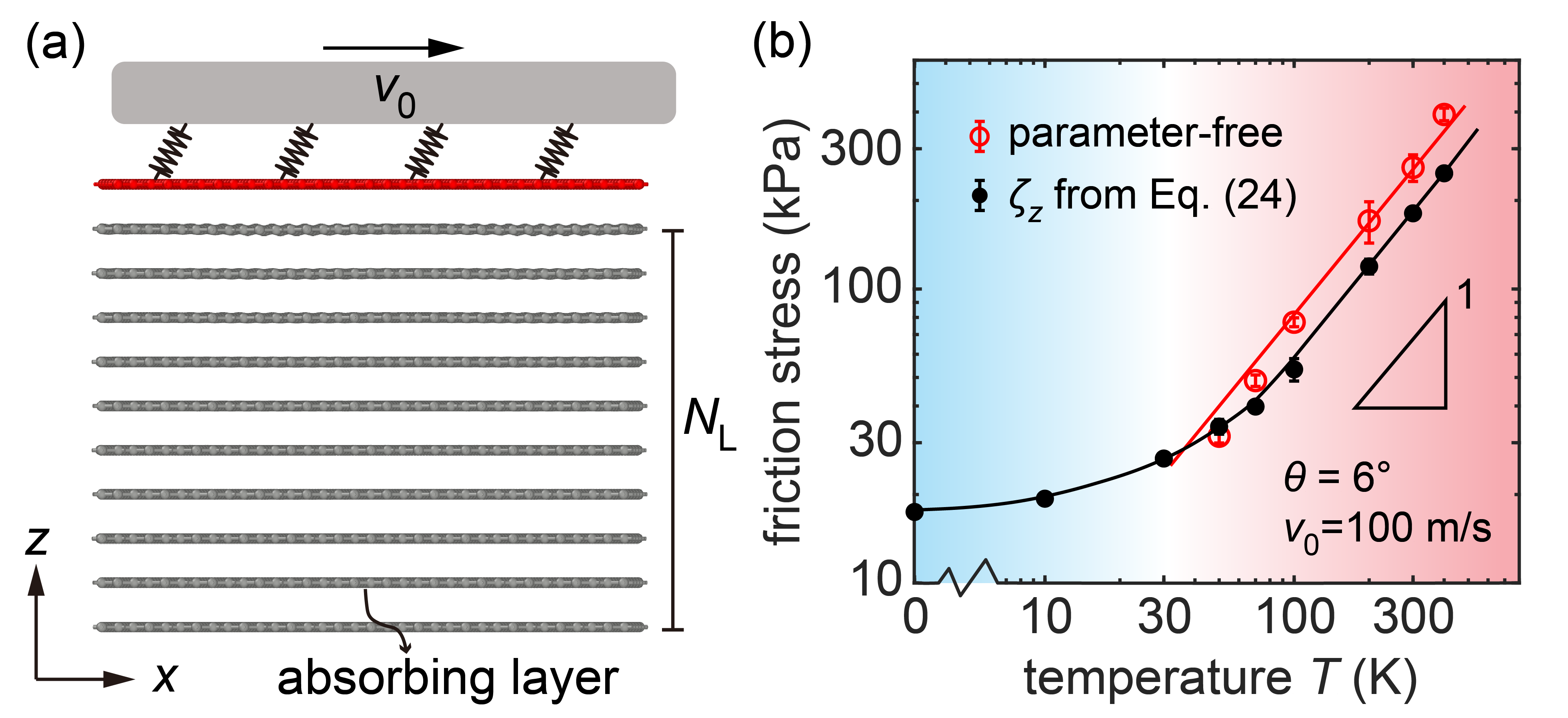}
\caption{Parameter-free simulation. (a) Schematic diagram of the model and set-ups. (b) Comparison of friction from parameter-free twisted graphene/multilayer graphite simulations (red) and effective bilayer graphene simulations with the theoretical damping $\zeta_z$ from Eq.~(24) (black). The solid line is guide to the eye.}
\label{fig:6}
\end{figure}

To obtain realistic kinetic friction in SSL systems, we built a multi-layer simulation model as shown in Fig.~\ref{fig:6}a, and adopted the parameter-free variational method -- applying damping to a far away boundary layer to correctly absorb phonons generated at the sliding interface \cite{Benassi.prb.2010}. The optimal boundary damping minimizes the back reflected energy by that remote boundary, and the corresponding maximal friction force is an approximation to the real friction that can be made as accurate as desired by increasing the layer number.\\

The simulation model consists of one layer of graphene slider and $N_L$ layers of Bernal graphite substrate. In our current simulation we use $N_L=10$, but we verified in a few test cases that the results are nearly same as $N_L=20$, once temperatures are not too low. PBCs are applied to $x$ and $y$ directions. The twist angle between the slider and the substrate is $\theta=6^\circ$, and the inter/intra-layer force field and the sliding set-ups are same as that in sect 6.1.
Perpendicular ``confining" springs
($k_z=0.33$~N/m) are attached to each slider atom  to limit the out-of-plane deformation. It can be expected that the deformation of the slider will increase for smaller $k_z$, resulting in higher sliding friction. The limited effects of $k_z$ on the out-of-plane deformation and friction is discussed in SI.
The center-of-mass of the slider is connected to the driving stage by a horizontal dragging spring ($K_p=100$~N/m), moving with a constant velocity $v_0$ along $x$.
The bottom layer of the substrate is fixed. The next layer up is connected to a Langevin thermostat with temperature $T$ and a boundary damping $\zeta_{N_L}$, which is the parameter that is variationally optimized by minimizing the backreflected phonon energy \cite{Benassi.prb.2010}.
The kinetic friction in each simulation is calculated from the time-averaged lateral force during a 10-ns steady sliding, multiple independent simulations are performed to obtain the converged value. \\

Before coming to the results, we should note that results are more and more reliable the higher the temperature, whereas for lower temperatures the multilayer thickness becomes insufficient and the optimal variational boundary damping parameter $\zeta_{N_L}$ increases to the maximum tolerable value, until optimization is lost, reflecting the excessive increase of phonon mean-free-path. That makes the multilayer variational approach essentially a high temperature one, not for practical use below $T_c$ where friction saturates.

Simulation results for friction in the range $T=50\sim400$~K are shown in Fig.~\ref{fig:6}b. There is good qualitative agreements between damping-based bilayer simulations with $\zeta_z$ from Eq.~(24), and parameter-free multilayer simulations. The friction stress at each temperature is of similar magnitude and both scale linearly with temperature (note that here $T > T_c$). Quantitatively, we find that with the theoretical $\zeta_z \sim 0.05~\mathrm{ps}^{-1}$ of Eq. (24), the kinetic friction is underestimated by a factor $\sim2$ with respect to the reference simulation, considered to be reliable.
Thus, for our particular exemplification of 2D materials-based SSL, the recommended damping coefficient of the effective bilayer should have been in the order of $0.1~\mathrm{ps}^{-1}$. Given the simplifications and approximations we employ, particularly the harmonic phonon assumption, this discrepancy is not surprising and not at all fatal, finally justifying the use with good theoretical control of an effective bilayer with damping.
It permits to estimate the friction stress for a SSL system at all temperatures above a very low $T_q \sim$ 15 K, at any velocity. With a typical experimental velocity ($v_0\sim 1~\mu\mathrm{m/s}$), we predict $\tau \sim 10^{-6}$~kPa -- an utterly negligible frictional stress compared with current experimental values for graphene, graphite and other 2D sliders, typically $1 \sim 100$~kPa. This huge gap confirms, as already suggested by area and velocity less than linear friction, that real, finite-size experimental contacts \textit{are not strictly structurally superlubric}, and the friction is instead dominated by the edges of the slider and/or defects at the interface \cite{Wang.rmp.2022}.

\newpage
\section{Discussion and Conclusions}

We presented analytical predictions for the kinetic friction of structurally superlubric 2D material interfaces, accompanied by exemplificative sliding simulations of twisted graphene bilayer, and of twisted graphene on a  semi-infinite bulk graphite substrate.
At low temperatures, we analytically  derive from basic Langevin formulations an effective damping coefficient that yields at low temperatures a bilayer sliding friction that equals that on a semi-infinite bulk.
At high temperatures, kinetic friction is directly obtained by fluctuation-dissipation; and an overall formula is proposed covering all temperatures. These analytical results compare very well with the simulated kinetic friction of the bilayer, once the theoretically obtained Langevin damping is used.
Finally, the equivalence of the theoretical effective bilayer to
a realistic 2D monolayer sliding on a semi-infinite bulk is validated by variational multilayer simulations.
The theoretical and realistic kinetic frictions agree to within a factor 2, which can be considered a very good result in view of the harmonic approximations used, and of the lack of adjustable parameters. Numerical values of the frictional stress obtained and validated, of about $10^{-6}$~kPa for a realistic sliding velocity $v_0\sim 1~\mu\mathrm{m/s}$ provide the very first quantitative measure of the tiny Stokes frictional dissipation connected with the surfing of the gossamer-like moir\'e pattern at an incommensurate 2D material interface. \\

Hard to detect as this predicted ultra-low friction clearly is, the actual unveiling of its nature and its value in a real case as presented here is nonetheless important on several accounts. Firstly, it physically formulates the problem, leading to clear, parameter-free results. Second, these results translate into numbers and temperature dependencies that were so far unknown. Third, they provide an element of clarity, showing that much of the supposedly superlubric kinetic friction reported in literature, as much as six orders of magnitude larger and with improper nonlinear velocity and area dependence, must be of different origin than SSL, being probably connected with stick-slip caused by the presence of edges, defects, and third bodies
-- whose eventual mitigation provides the directions for the future realization of superlubricity.

\newpage
\section*{Acknowledgments}

E.T. and J.W. acknowledge support from ERC ULTRADISS Contract No. 834402. Support by the Italian Ministry of University and Research through PRIN UTFROM N. 20178PZCB5 is also acknowledged.
M.M. acknowledges the financial support from the National Natural Science Foundation of China (No. 11890673 and 51961145304).
J.W. acknowledges the computing resources support from National Supercomputer Center in Tianjin.
We are grateful for  discussions with A. Khosravi, A. Silva, and A. Vanossi.


\bibliographystyle{unsrt}
\bibliography{ref}

\end{document}



\title{Supplementary Information \\
Kinetic Friction of Structurally Superlubric 2D Material Interfaces}

\author{Jin Wang}
\affiliation{International School for Advanced Studies (SISSA), I-34136 Trieste, Italy}
\affiliation{International Centre for Theoretical Physics, I-34151 Trieste, Italy}

\author{Ming Ma}
\affiliation{State Key Laboratory of Tribology, Department of Mechanical Engineering, Tsinghua University, Beijing 100084, China}

\author{Erio Tosatti}
\email{tosatti@sissa.it}
\affiliation{International School for Advanced Studies (SISSA), I-34136 Trieste, Italy}
\affiliation{International Centre for Theoretical Physics, I-34151 Trieste, Italy}
\affiliation{CNR-IOM, Consiglio Nazionale delle Ricerche - Istituto Officina dei Materiali, c/o SISSA,  34136, Trieste, Italy}


\maketitle
\tableofcontents
\newpage

\section{Methods for structural optimization}

The energy optimization is performed with open-source code LAMMPS \cite{Plimpton.jcp.1995,Thompson.compphyscomm.2022}. The bilayer simulation models with twist angle $\theta$ ranges from $0.3^\circ$ to $30^\circ$ are created with periodic boundary conditions (PBC) along $x$ and $y$ directions. The interlayer and intralayer interaction is described by registry-dependent interlayer potential (ILP) and REBO force field respectively \cite{Ouyang.nanolett.2018, Brenner.2002}. The carbon atom in each layer is tethered by a linear $z$-directional spring to its original position to mimic the normal direction elasticity. The spring constant, which reproduces the moir\'e height in twisted bulk graphite, is $k_z=0.33$~N/m. During the structural optimization, the in-plane stress is kept to be zero, $p_{xx}=p_{yy}=p_{xy}=0$. The optimization is performed by FIRE \cite{Bitzek.prl.2006} together with CG algorithms with several loops. The convergence criterion is when the largest single atom force $F_i < 10^{-6}~\mathrm{eV/\AA}$.

\section{Negligible effects from in-plane damping}
The in-plane damping $\zeta_x$ (and $\zeta_y$) is negligible (compared to the out-of-plane $\zeta_z$) in the damping-based bilayer graphene simulations. This has been realized before (e.g., Ref.~\cite{Persson.2000}) and can be understood here by defining an anisotropy factor:
\begin{equation}
    r_{aniso} = \frac{\zeta_z}{\zeta_x} = \frac{|\Phi_{12}^{zz}|^2}{|\Phi_{12}^{xx}|^2}
\end{equation}
The $z$-direction force constant is $\Phi_{12}^{zz}=2.7$~N/m (could also be estimated from elastic constant $C_{33}=38.7$~GPa by $\Phi_{12}^{zz}=C_{33} A_\mathrm{C}/d_0$), and the force constant $\Phi_{12}^{xx}$ could be estimated from the interlayer shearing $C_{44}=5.0$~GPa \cite{Bosak.prb.2007}. Since $\Phi_{12}^{zz}/\Phi_{12}^{xx}=C_{33}/C_{44}$, the anisotropy factor is in the order of $10^2$ -- the friction contribution from in-plane damping is thus negligible.

\newpage
\section{Damping coefficient from anisotropic Debye model}
Considering an anisotropic Debye model with dispersion relation \cite{Chen.prb.2013}:
\begin{equation}
    \omega^2=v_\mathrm{in}^2 (q_x^2+q_y^2) + v_\mathrm{out}^2 q_z^2
\end{equation}
where $v_\mathrm{in}$ and $v_\mathrm{out}$ is the in-plane and out-of-plane sound speed. The first Brillouin zone of this model is assumed to be an ellipsoid:
\begin{equation}
    (\frac{q_x}{q_{x0}})^2+(\frac{q_y}{q_{y0}})^2+(\frac{q_z}{q_{z0}})^2=1
\end{equation}
/here for 2D materials, we take $q_{x0}=q_{y0}$. The Debye frequencies along in-plane and out-of-plane directions could be defined as $\omega_{Dx}=\omega_{Dy}=v_\mathrm{in} q_{x0}$ and $\omega_{Dz}=v_\mathrm{out} q_{z0}$.

The density of state is \cite{Kittel.1996}:
\begin{equation}
\begin{aligned}
    \rho(\omega)&=\frac{V}{8\pi^3} \int \frac{\mathrm{d}S}{v_g} \\
    &=2\times \frac{V}{8\pi^3} \iint \frac{1}{v_g} \sqrt{1+(\frac{\partial q_z}{\partial q_x})^2+(\frac{\partial q_z}{\partial q_y})^2} \mathrm{d}q_x \mathrm{d}q_y
\end{aligned}
\end{equation}
where $V$ is the volume of the unit cell, $v_g=|\nabla_q \omega|$ is the magnitude of the group velocity of a phonon, $S$ represents the surface ``area'' of the zone boundary. By implementing the polar coordinate substitution $q_x=r \cos\varphi$ and $q_y=r \sin\varphi$ ($r \in [0, \omega/v_\mathrm{in}], \varphi \in [0, 2\pi]$.), one could simplify the above equation to
\begin{equation}
    \rho(\omega) = \frac{V \omega^2}{2\pi^2 v_\mathrm{in}^2 v_\mathrm{out}}
\end{equation}
This result holds for $\omega<\omega_{Dz}$ -- consistent with our work. Substituting Eq. (S5) back into Eq. 23 in the maintext, we could get the damping coefficient:
\begin{equation}
    \zeta_z = \frac{|\Phi_{12}^{zz}|^2}{m^2} \frac{V}{4\pi v_\mathrm{in}^2 v_\mathrm{out}}
\end{equation}
By using sound speed $v_\mathrm{in}=22~\mathrm{km/s}$ and $v_\mathrm{out}=1.48~\mathrm{km/s}$ \cite{Bosak.prb.2007}, one could get the damping coefficient $\zeta \approx 0.02~\mathrm{ps^{-1}}$ -- agree qualitatively with the estimation given in the main text.

\newpage

\section{Thermal fluctuations from simulations}
In this section, we give details on the mean-square (out-of-plane) thermal fluctuation
$\langle H_T^2 \rangle$
of the interfacial layer (region B) and its temperature dependence. 

For bilayer graphene simulations (described in maintext Sect.~6.1), we can get trajectory of all atoms, $x_i(t)$, $y_i(t)$, and $z_i(t)$.
According to definition,
\begin{equation}
    \langle H_T^2 \rangle = \sum_q \langle |h_q|^2 \rangle
\end{equation}
where $\langle |h_q|^2 \rangle$ could be estimated from 2D Fourier transform of $z(x,y)$:
\begin{equation}
    \langle |h_q|^2 \rangle = \langle |\mathrm{FFT}_q|^2 \rangle
\end{equation}
here $|...|$ represents complex modulus, and
\begin{equation}
    \mathrm{FFT}(q_x, q_y)= \frac{4}{N_x N_y}\sum_{j=0}^{N_x-1} \sum_{k=0}^{N_y-1} 
    \exp(- \frac{2\pi i q_x j}{N_x}) \exp(-\frac{2\pi i q_y k}{N_y}) z(x_j,y_k)
\end{equation}
where $z(x_j,y_k)$ is the out-of-plane position of the substrate (or slider), remapping from the original honeycomb lattice to square lattice with spatial resolution $N_x\times N_y$.

\begin{figure}[ht!]
\centering
\includegraphics[width=0.88\linewidth]{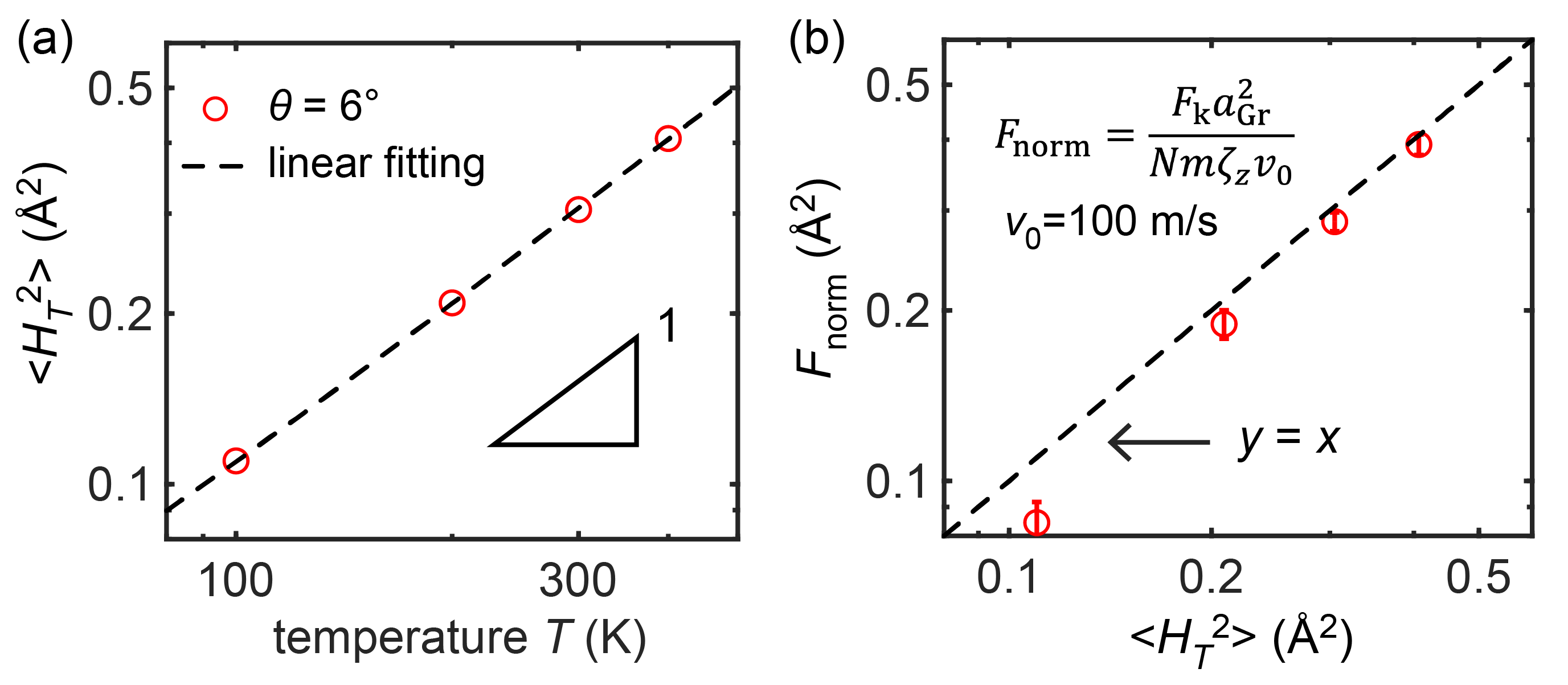}
\caption{(a) Temperature dependence of mean-square (out-of-plane) thermal fluctuations $\langle H_T^2 \rangle$. (b) A direct comparison between the ``normalized friction'' $F_\mathrm{norm}$ and $\langle H_T^2 \rangle$. Slope with $y=x$ is plotted as a guide for the eye.}
\label{fig:S1}
\end{figure}

The temperature dependence of $\langle H_T^2 \rangle$ (for $T>T_c$) is shown in Fig.~\ref{fig:S1}a.
Note that at low temperature limit ($T\to0$), the mean-square-corrugation becomes $H_\mathrm{moire}^2$. For $\theta=6^{\circ}$, it is approximately $10^{-2}~\mathrm{\AA^2}$.
From simulation results, we can determine the value of $c_2$ by:
\begin{equation}
    c_2= \frac{\kappa \langle H_T^2 \rangle} {k_\mathrm{B} T a_\mathrm{Gr}^2} \approx2.8
\end{equation}
where $\kappa$ is the bending stiffness of monolayer graphene, $a_\mathrm{Gr}$ is the lattice constant of graphene.
Substituting this $c_2$ back into Eq.~(28) in the maintext, we find that the friction force estimated from our theory is in good agreement with the simulation results (Fig.~5d in the maintext).

The origin of this temperature dependence of friction (at high temperatures $T>T_c$) is $\langle H_T^2 \rangle$, as we formulate in Eq.~(27) in the maintext. This could also be demonstrated directly from our simulation as shown in Fig.~\ref{fig:S1}b. The errorbar of the friction force is the standard deviation of three independent simulations.
Simulation results show that $F_\mathrm{norm}=F_\mathrm{k}a_\mathrm{Gr}^2 (N m \zeta_z v_0)^{-1} \simeq \langle H_T^2 \rangle$ at high temperatures, which lead immediately to:
\begin{equation}
    F_\mathrm{k} \simeq
     \frac{N m \zeta_z \langle H_T^2 \rangle v_0}{a_\mathrm{Gr}^2}
\end{equation}\\

\section{Corrugations with different $k_z$}

In the maintext, we use the same out-of-plane restriction $k_z$ to slider and substrate. 
This set-up naturally gives rise to the same moir\'e height and thermal corrugations. In real nature there are cases where $k_z$ is different for slider and substrate, e.g, graphite/hBN heterostructures, twisted monolayer graphene on Bernal graphite, etc. Discussions on fluctuations and sliding frictions for these variety of systems with ``asymmetric'' $k_z$ will be given in this section.

\begin{figure}[ht!]
\centering
\includegraphics[width=0.5\linewidth]{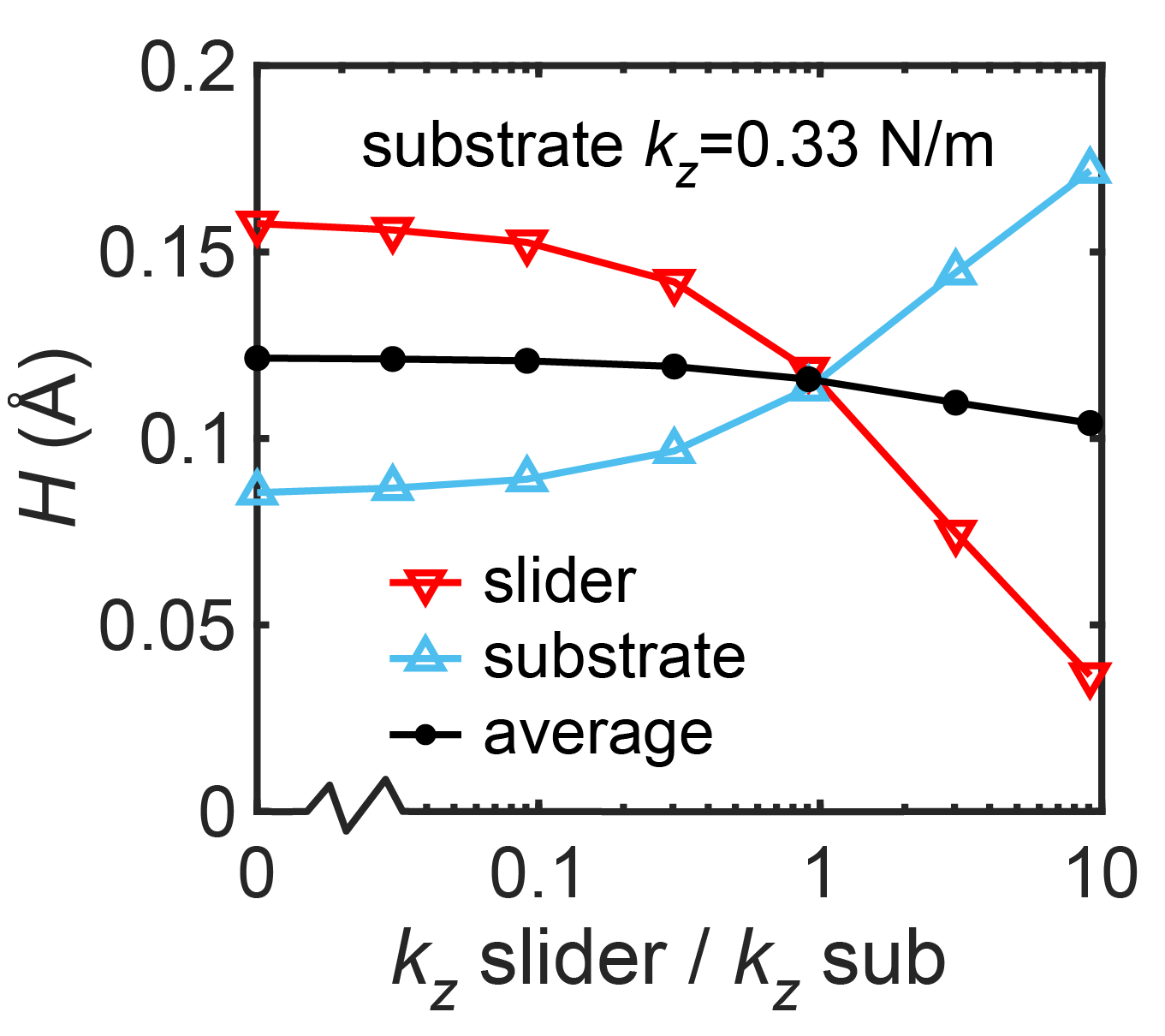}
\caption{Dependence of moir\'e corrugation on the ratio of slider to substrate $k_z$. The spring constant for the substrate is fixed to be $k_z=0.33$~N/m.}
\label{fig:S2}
\end{figure}

At low temperature limit, the sliding friction is dominated by the moir\'e corrugation, i.e., Eq.~(12) in the maintext.
With different $k_z$ for the substrate and slider, the moir\'e corrugation changes correspondingly.
Thus, Eq.~(12) could be generalized by defining $H_\mathrm{sub}$ and $H_\mathrm{sli}$ whose values can be determined by finding the global minimum potential energy.
Here, instead, we show the simulation results of moir\'e corrugation for different slider's $k_z$.
Adopting the same optimization protocol with substrate's $k_z=0.33$~N/m and slider's $k_z$ ranging from $0$ to $3~\mathrm{N/m}$, the moir\'e corrugations for the slider $H_\mathrm{sli}$ and substrate $H_\mathrm{sub}$ of a test case with $6^\circ$-twist bilayer graphene are shown in Fig.~\ref{fig:S2}.
From simulations, whether $k_z$ of the slider is zero or equals to that of the substrate has weak influence on $H_\mathrm{sub}$ and negligible effect on the average moir\'e height $(H_\mathrm{sub}+H_\mathrm{sli})/2$.
Therefore, it is safe to continue using Eq.~(12) in the maintext to approximate the moir\'e corrugations for more general SSL systems.

\begin{figure}[ht!]
\centering
\includegraphics[width=0.5\linewidth]{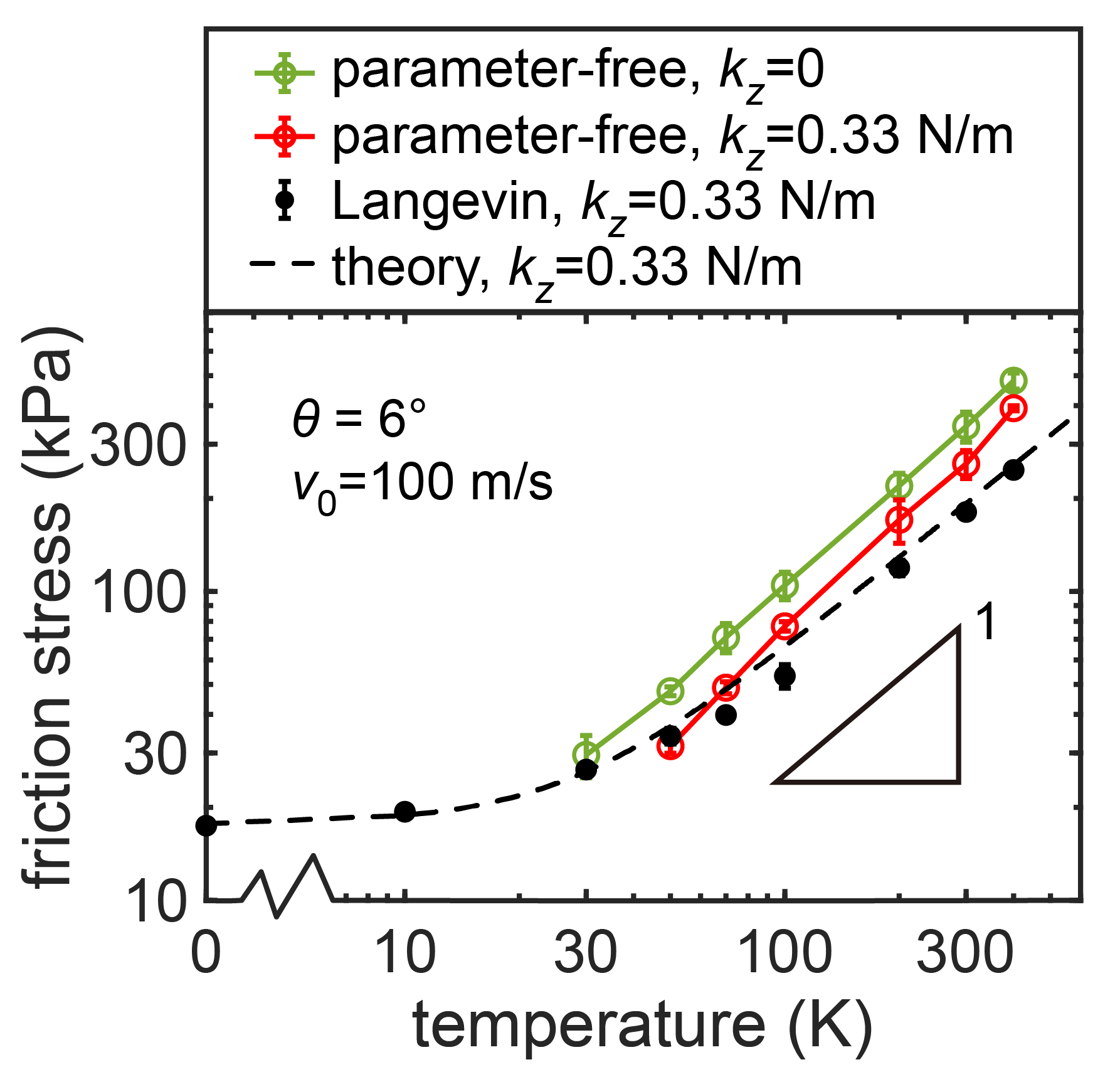}
\caption{Temperature dependence of frictional stress. Green and red circles: parameter-free simulation results for slider's $k_z$ equal to $0$ and $0.33$~N/m. Black dots: Langevin-based simulation results (with $\zeta_z$ estimated from Eq.~24 in the maintext) for $k_z^\mathrm{sli}=k_z^\mathrm{sub}=0.33$~N/m. Dashed line: theoretical estimation with $k_z^\mathrm{sli}=k_z^\mathrm{sub}=0.33$~N/m.}
\label{fig:S3}
\end{figure}

For high temperature cases, where moir\'e fluctuations become negligible, the sliding friction is dominated by mean-square-fluctuations $\langle H_T^2 \rangle$.
The larger $k_z$ results in higher deformation energy, which leads to a decrease in $\langle H_T^2 \rangle$ (and the sliding friction) at the same temperature.
We test on the parameter-free multilayer simulation system with two cases: $k_z$ of the slider is equal to $0$ and $0.33$~N/m.
From the simulation results shown in Fig.~\ref{fig:S3}, the frictional stress at room temperature for $k_z=0$ case (green) increased by $\approx30\%$ compared to the finite $k_z$ case (red). Simulation and theoretical results for the bilayer system with equal $k_z$ are also shown in the figure.

\newpage
\bibliographystyle{unsrt}
\bibliography{ref}